\documentclass[submission,copyright,creativecommons]{eptcs}
\providecommand{\event}{MARS 2018} 
\usepackage{breakurl}             
\usepackage{underscore}           

\usepackage{amssymb}
\usepackage{graphicx}
\graphicspath{{./eps/}}
\DeclareGraphicsExtensions{.eps}
\usepackage{subfigure}
\usepackage{wrapfig}
\usepackage{amsmath,amsthm}
\usepackage{extarrows}
\usepackage{uppaal}
\usepackage{multirow}
\usepackage{multicol}

\title{A Modeling Framework for Schedulability Analysis of Distributed Avionics Systems}
\author{Pujie Han \qquad Zhengjun Zhai
\institute{School of Computer Science and Engineering\\
Northwestern Polytechnical University\\
Xi'an, China}
\email{\{hanpujie,zhaizjun\}@mail.nwpu.edu.cn}
\and
Brian Nielsen \qquad Ulrik Nyman
\institute{Department of Computer Science\\
Aalborg University\\
Aalborg, Denmark}
\email{\quad \{bnielsen,ulrik\}@cs.aau.dk}
}
\def\titlerunning{A Modeling Framework for Schedulability Analysis of Distributed Avionics Systems}
\def\authorrunning{P. Han, Z. Zhai B. Nielsen \& U. Nyman}

\begin{document}

\maketitle

\begin{abstract}
This paper presents a modeling framework for schedulability analysis of distributed integrated modular avionics (DIMA) systems that consist of spatially distributed ARINC-653 modules connected by a unified AFDX network. We model a DIMA system as a set of stopwatch automata (SWA) in \uppaal to analyze its schedulability by classical model checking (MC) and statistical model checking (SMC). The framework has been designed to enable three types of analysis: global SMC, global MC, and compositional MC. This allows an effective methodology including (1) quick schedulability falsification using global SMC analysis, (2) direct schedulability proofs using global MC analysis in simple cases, and (3) strict schedulability proofs using compositional MC analysis for larger state space. The framework is applied to the analysis of a concrete DIMA system.
\end{abstract}

\section{Introduction}\label{sec:intro}

In the avionics industry, Distributed Integrated Modular Avionics (DIMA) has been widely recognized as a promising architecture and the next generation of Integrated Modular Avionics (IMA). A DIMA system installs standardized IMA modules in spatially distributed processors\cite{wang2013research} that communicate through a unified bus system\cite{annighofer2014systems} such as an AFDX network. Avionics functions residing on the IMA modules are implemented in the form of application software running in an ARINC-653\cite{arinc653} compliant operating system. The generic distributed structure of DIMA significantly improves performance and reliability as well as lowers weight and cost, while it also dramatically increases the complexity of schedulability analysis. A schedulable DIMA system should fulfil not only the temporal requirements of real-time tasks in each ARINC-653 module but also several communication constraints among the distributed nodes. As a result, the DIMA architecture requires the system integrators to analyze schedulability considering both computation and communication.

The development of model checking based approaches has currently become an attractive topic for the schedulability analysis of complex real-time systems due to the sufficient expressiveness of formal models. The techniques of classical model checking (MC) describe schedulability as temporal logic properties and verify the properties via symbolic state space exploration. Unfortunately, when being applied to a complete avionics system, all of them suffer from an inevitable problem of state space explosion, which makes the exact symbolic model checking practically infeasible.

Accordingly, Statistical Model Checking (SMC) is proposed as a promising technique that has powerful facilities of formal modeling as well as avoids the state-space explosion of classical model checking. A SMC engine runs and monitors a number of simulation processes, quickly estimating the statistical results of the satisfaction or violation of certain properties. However, SMC cannot provide any guarantee of schedulability but quick falsification owing to its nature of statistical testing. Therefore, it is reasonable to apply both classical and statistical model checking to the schedulability analysis of avionics systems.

{\small\textbf{Related work:}} We found no studies that analyzed the schedulability of distributed avionics systems as a whole including the network by model checking. The related research isolates computation modules from their underlying network, thereby considering these nodes as independent hierarchical scheduling systems or investigating the network in isolation, which possibly leads to pessimistic results. There have been works using model-checking approaches to analyze the temporal behavior of individual avionics modules in various formal models such as Coloured Petri Nets (CPN)\cite{dodd2006coloured}, preemptive Time Petri Nets (pTPN)\cite{carnevali2011formal}, Timed Automata (TA)\cite{amnell2003times}, and StopWatch Automata (SWA)\cite{cicirelli2012development}, and verify schedulability properties via state space exploration. For hierarchical scheduling systems, some studies\cite{carnevali2013compositional,sun2014component,boudjadar2014compositional} exploit the inherent temporal isolation of ARINC-653 partitions\cite{arinc653} and analyze each partition separately, but they ignore the behavior of the underlying network or the interactions among partitions. Thus these methods are not applicable to DIMA environments in which multiple distributed ARINC-653 partitions communicate through a shared network to perform an avionics function together.

{\small\textbf{Contributions:}} In this paper, we present a modeling framework for schedulability analysis of DIMA systems that are implemented as a set of \uppaal SWA, i.e. the TA extended with stopwatches\cite{cassez2000impressive} in \uppaal. The framework combines compositional and global analysis by classical and statistical model checking. The main contributions of this paper are summarized as follows:
\begin{itemize}
  \item \emph{Modeling of DIMA systems} covers the major behavior of two-level ARINC-653 compliant schedulers, periodic/sporadic tasks, intra-partition synchronization, and inter-partition communications through an AFDX network.
  \item \emph{Compositional analysis using classical model checking} verifies the model of each ARINC-653 partition including its environment individually and then assemble the local results together to derive conclusions about the schedulability of an entire system.
  \item \emph{Global analysis using statistical model checking} allows users to quickly falsify non-schedulable configurations by SMC hypothesis testing, which can handle a complete system model and avoid an exhaustive exploration of the state space.
\end{itemize}

The rest of the paper is organized as follows. Section \ref{sec:avionics} describes the structure of a DIMA system. An overview of the modeling framework is presented in section \ref{sec:modelframework}, where the methods of compositional and global analysis are briefly outlined. In section \ref{sec:modeldetails}, we detail the \uppaal models of the framework. Section \ref{sec:experiment} shows an experiment on a concrete DIMA system, and section \ref{sec:conclusion} finally concludes.

\section{Avionics System Description}\label{sec:avionics}

\begin{figure*}[!t]
\begin{minipage}{.5\textwidth}
\centering
\includegraphics[width=2.0in]{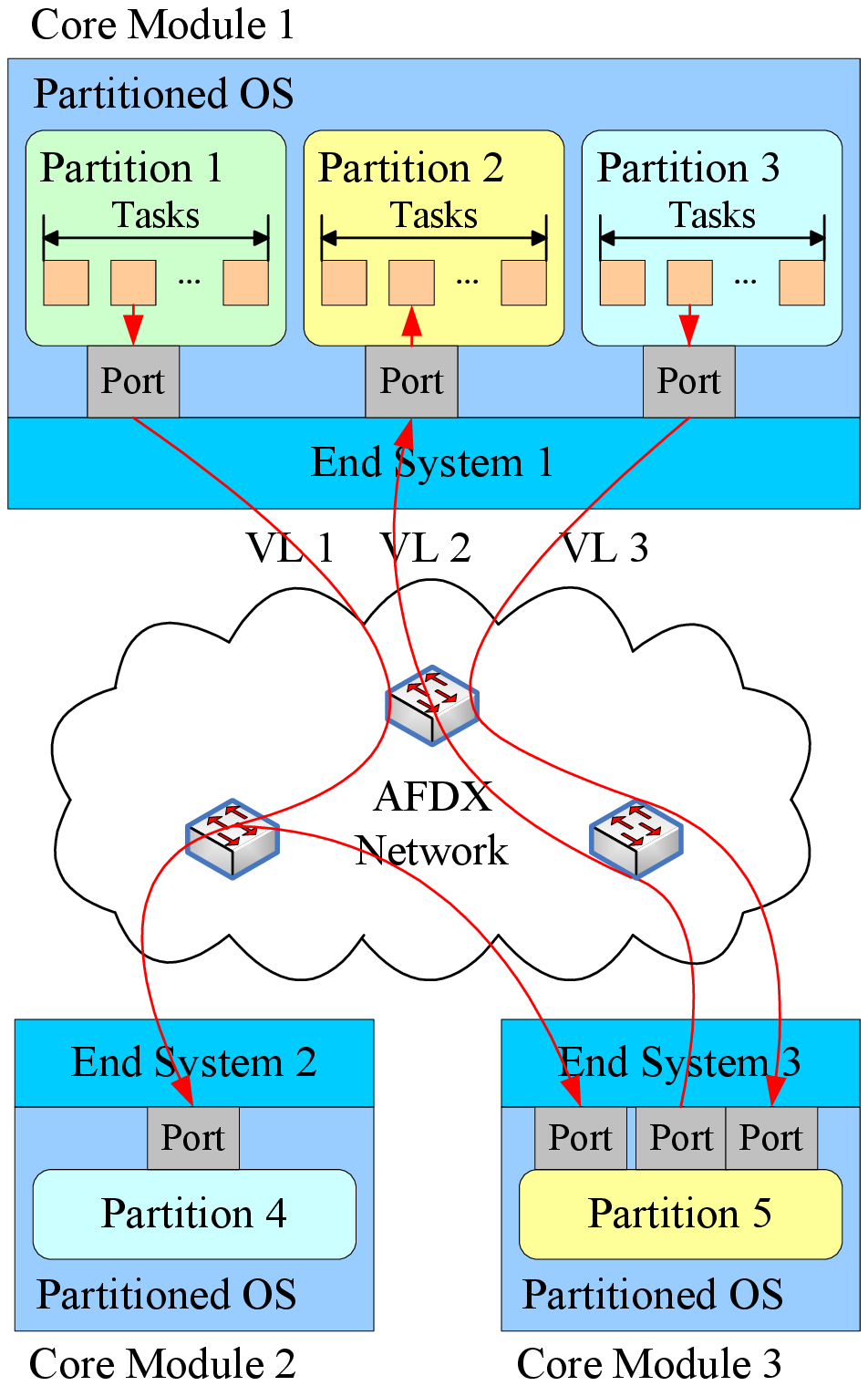}
\caption{An Example of DIMA systems}
\label{fig:afdx}
\end{minipage}%
\begin{minipage}{.5\textwidth}
\centering
\vspace{19pt}
\includegraphics[width=2.3in]{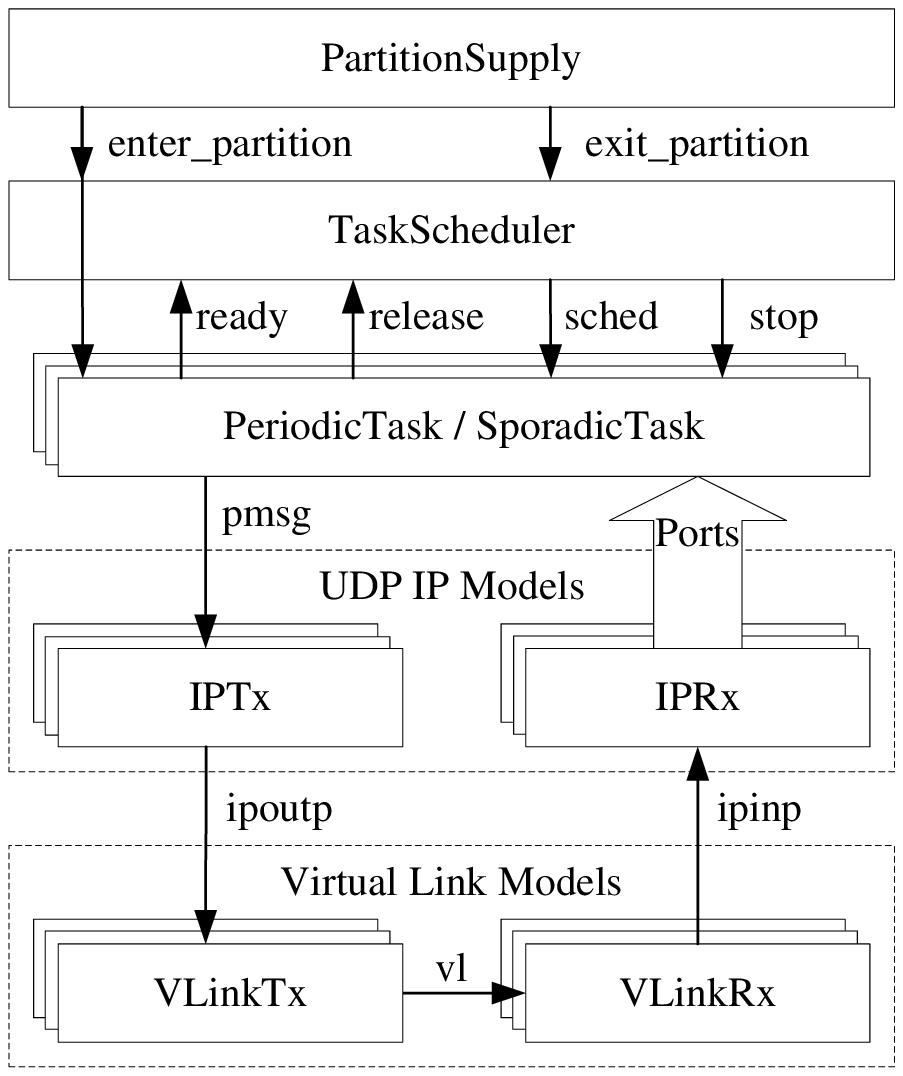}
\vspace{15pt}
\caption{An Overview of Modeling Framework}
\end{minipage}%
\label{fig:overview}
\end{figure*}

We consider a generic architecture of a DIMA system with several ARINC-653 modules connected by an AFDX network shown in Fig.\ref{fig:afdx}. There is a three-layer structure in the DIMA architecture consisting of scheduling, task, and communication layers.

The \emph{scheduling layer} comprises the scheduling facilities for generic computation resources in a DIMA system, where physically distributed modules with independent computational power execute application tasks simultaneously. The tasks run in a partitioned operating system which provides a two-level scheduling mechanism and achieves temporal isolation between ARINC-653 partitions. In such a scheduling system, partitions are scheduled by a Time Division Multiplexing (TDM) scheduler and each partition also has its local scheduling policy, preemptive Fixed Priority (FP), to handle the internal tasks\cite{arinc653}.

All the application tasks executing avionics functions constitute the \emph{task layer}. We consider a task as the smallest scheduling unit, each of which can be executed concurrently with other tasks in the same partition. Assume that jobs of each task are scheduled repeatedly. We define two task types: \emph{periodic tasks} and \emph{sporadic tasks}. A periodic task has a fixed release period, while a sporadic task is characterized by a minimum separation between consecutive jobs.

The \emph{communication layer} provides the services of inter-partition communication over a common AFDX network. The AFDX protocol stack realized by an End System(ES) interfaces with the task layer through ARINC-653 ports. Based on the AFDX protocol structure, the communication layer is further divided into UDP/IP layer and Virtual Link layer, where a Virtual Link (VL) ensures an upper bound on end-to-end delay.

The communication layer also affects the schedulability of the system. According to the ARINC-653 standard\cite{arinc653}, there are two types of ARINC-653 ports, sampling ports and queuing ports. A sampling port can accommodate at most a single message that remains until it is overwritten by a new message. Moreover, a refresh period is defined for each sampling port. This attribute provides a specified arrival rate of messages, regardless of the rate of read requests from tasks. In contrast, a queuing port is allowed to buffer multiple messages in a message queue with a fixed capacity. However, the operating system is not responsible for handling overflow from the message queue.

In our framework, we verify the three following schedulability properties of DIMA systems: (1) All the tasks meet their deadlines in each partition. (2) The refresh period of any sampling port is guaranteed. (3) The overflow from any queuing ports is avoided.

\section{An Overview of the Modeling Framework}\label{sec:modelframework}

\subsection{An Outline of the UPPAAL Models}\label{sec:modeloverview}

The \uppaal templates in the modeling framework are organized as the above layered structure. Fig.\ref{fig:overview} shows an overview of these templates together with the channels between them.

The scheduling layer consists of two TA templates \uppTemp{PartitionScheduler} and \uppTemp{TaskScheduler}. The \uppTemp{PartitionScheduler} model provides the service of TDM partition scheduler for any partition. The \uppTemp{Tas}-\\ \uppTemp{kScheduler} model implementing FP scheduling policy allocates processor time to the task layer only when the partition is active. Hence \uppTemp{PartitionScheduler} sends notification on the broadcast channels \uppSync{enter_partition} and \uppSync{exit_partition} to \uppTemp{TaskScheduler} when entering and leaving its partition, respectively.

The task layer contains a set of task models which are instantiated from two SWA templates \uppTemp{Periodi}-\\ \uppTemp{cTask} and \uppTemp{SporadicTask}. A task model describes an instance of an avionics program. Since the tasks in a partition are scheduled by a task scheduler, we define four channels \uppSync{ready}, \uppSync{release}, \uppSync{sched} and \uppSync{stop} as a set of scheduling commands to communicate between task templates and \uppTemp{TaskScheduler}. Moreover, the priority ceiling protocol is implemented by mutexes in task models to deal with intra-partition synchronization.

The communication layer comprises two types of models: UDP/IP and VL models. The UDP/IP models are divided into two TA templates \uppTemp{IPTx} and \uppTemp{IPRx}, which calculate the delivery latency of the UDP/IP layer in a transmitting ES and a receiving ES respectively. When sending a message to an ARINC-653 port, the source task notifies the destination \uppTemp{IPTx} via the broadcast channel \uppSync{pmsg}. In the link layer, two TA templates \uppTemp{VLinkTx} and \uppTemp{VLinkRx} model the total latency of a VL through the transmitting ES and the reception network respectively. The channel \uppSync{vl} connects \uppTemp{VLinkTx} and \uppTemp{VLinkRx} in the same VL. Additionally, there are also two broadcast channels \uppSync{ipoutp} and \uppSync{ipinp} between the UDP/IP and VL models in opposite directions.

\subsection{Integration of MC and SMC}

In \uppaal SMC, a model comprises networks of Stochastic Timed Automata (STA), which is designed as a stochastic interpretation and extension of the TA formalism of \uppaal classic\cite{david2015uppaal}. To integrate SMC into a common framework, we adapt the above templates for STA with the following features:
\begin{itemize}
\item\emph{Input-enabledness}: Only broadcast channels, which can attach to an arbitrary number of receivers, do we use in the modeling framework to ensure the input-enabled property that no input actions are prevented from being sent to a STA.
\item\emph{Deterministic bounded delays}: The semantics of non-deterministic delays in TA is replaced with probability distributions. The bounded delay at a location of templates is defined as a uniform distribution in STA. For example, \uppTemp{PeriodicTask} executes computing operations at a location \uppLoc{Running} with a bounded delay interval $[\mathit{bcet},\mathit{wcet}]$. Thus we assume this execution time to be random samples from the uniform distribution $U(\mathit{bcet},\mathit{wcet})$.
\item\emph{Deterministic unbounded delays}: The unbounded delays are interpreted as exponential distributions in STA. We take the model \uppTemp{SporadicTask} for example. There is a minimum separation but no maximum constraint between consecutive releases of a sporadic task. We describe this unbounded separation as an exponential distribution with an empirical rate at a location \uppLoc{WaitNextRelease}.
\item\emph{Non-zenoness}: We adopt two modeling principles to reduce the risk of zenoness in \uppaal: (1) Avoid the loops composed of permanently enabled edges and urgent/committed locations where time cannot progress. (2) Provide normal locations with two types of outgoing edges, which are either event-driven edges that contain input channels or time-driven edges that do not have input labels but guards making time progress at source locations. Since \uppaal SMC can detect zeno runs\cite{david2015uppaal}, the modeling framework has been checked thoroughly to achieve non-zenoness.
\end{itemize}

\subsection{Schedulability Analysis by MC and SMC}

\begin{figure}[!t]
\centering
\includegraphics[width=4in]{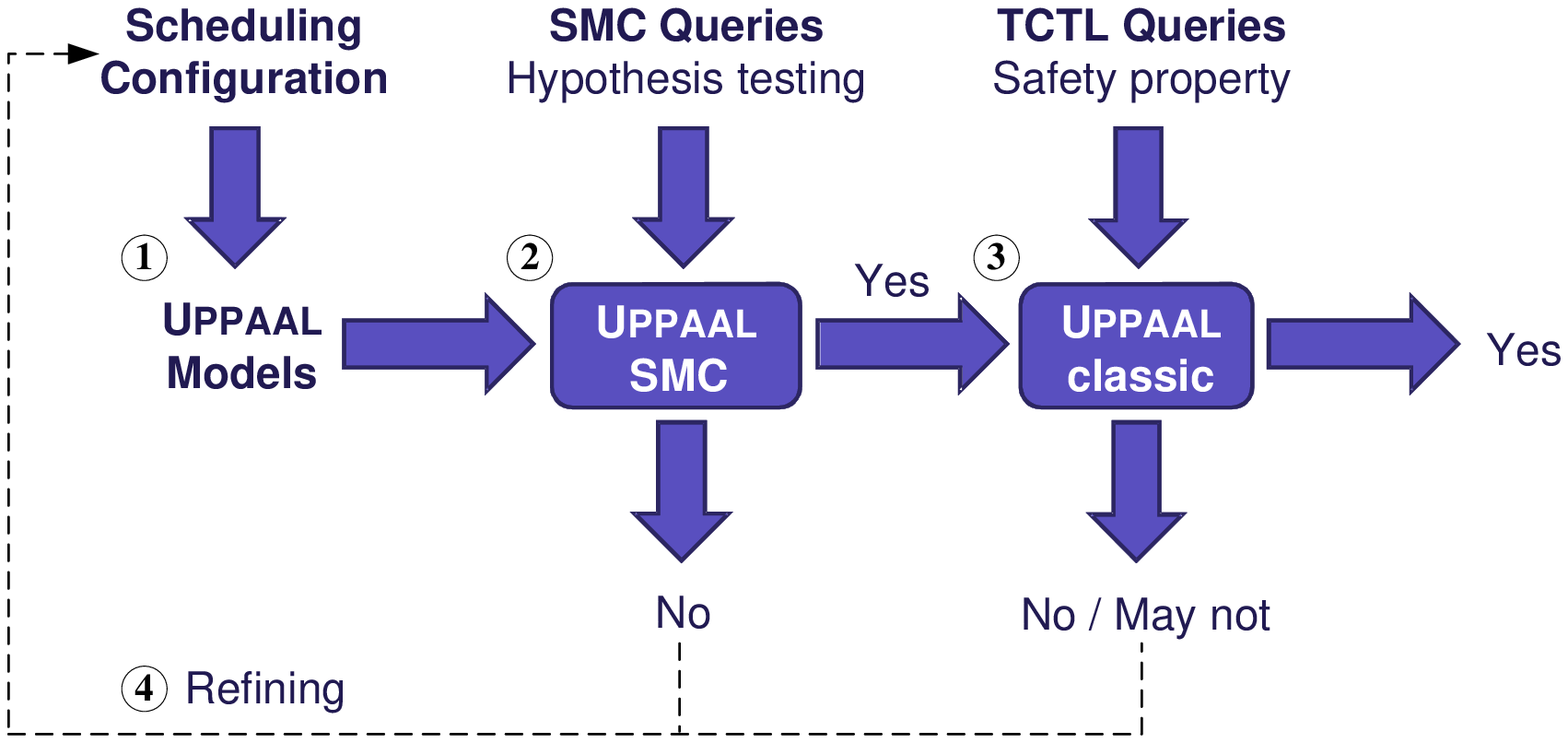}
\caption{The Procedure for Schedulability Analysis}
\label{fig:procd}
\end{figure}

On the basis of the above models, we introduce the procedure for our schedulability analysis, which combines classical and statistical model checking together. Fig.\ref{fig:procd} shows the four steps in the procedure: (1) Scheduling configuration is encoded into the \uppaal model as a structure array. (2) We perform hypothesis testing of SMC for the model to falsify non-schedulable configuration rapidly. (3) If the model goes through the SMC test, its schedulability should be verified by classical symbolic MC. (4) We refine the configuration that fails steps (2,3) and restart step (1).

When we apply classical MC to the analysis of a DIMA system, the schedulability constraints are expressed and verified as a safety property of SWA models. We add a boolean variable $\uppVar{error}$ with the initial value $\uppConst{False}$ to \uppaal templates for this purpose. Once the schedulability is violated, the related model will assign the value $\uppConst{True}$ to $\uppVar{error}$ immediately. Thus, the schedulability is replaced with this safety property $\varphi$:
\begin{equation}\label{eq:safetylogic}
\uppPropAG{not error},
\end{equation}
which belongs to a simplified subset of TCTL used in \uppaal.

According to the size of state space, we choose either a global or compositional MC analysis. The system models with small size can be handled by the \emph{global analysis} where the modeling elements of all the partitions in the system are instantiated and checked directly. Nevertheless, most concrete system models have larger state space, thereby making the global analysis infeasible. To reduce the state space in this case, we perform a \emph{compositional analysis} which check each partition including its environment individually. A set of message interface automata is built to model the environment for a partition.

The schedulability can be obtained from the satisfaction of $\varphi$, i.e. the MC result ``Yes'' in Fig.\ref{fig:procd}. However, since the symbolic MC of \uppaal for SWA introduces a slight over-approximation\cite{cassez2000impressive}, we cannot conclude non-schedulability from the MC results ``No'' or ``May not'' with certainty. Therefore, we derive non-schedulability from SMC testing rather than from the verification of $\varphi$.

Considering the scalability of SMC, we only use a global analysis in \uppaal SMC. The schedulability of a complete avionics system is described as following queries of hypothesis testing:
\begin{equation}\label{eq:schqry}
\uppProp{Pr[<=\ M](<> error)\ <=\ }\theta,
\end{equation}
where $M$ is the time bound on the simulations and $\theta$ is a very low probability. Since \uppaal SMC approximates the answer using simulation-based tests, we can falsify non-schedulable configuration (i.e. the SMC result ``No'' in Fig.\ref{fig:procd}) rapidly by finding counter-examples but identify schedulable ones only with high probability ($1-\theta$) (i.e. the SMC result ``Yes'' in Fig.\ref{fig:procd}). Hence, the configuration that goes through the SMC tests should be validated by symbolic MC to ensure the schedulability of the corresponding system.

\section{UPPAAL Models}\label{sec:modeldetails}

In this section, we detail the major \uppaal templates according to the layered structure from top down.

\subsection{Scheduling Layer Models}

\begin{figure}[!t]
\centering
\includegraphics[width=3.8in]{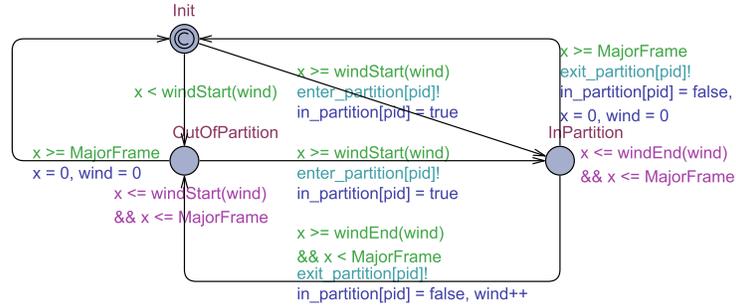}
\caption{PartitionScheduler Model}
\label{fig:ps}
\end{figure}

{\small\textbf{PartitionScheduler template}} In the scheduling layer, a partition is activated only during its partition windows within every major time frame $T_{mf}$. We build a TA model \uppTemp{PartitionScheduler}(See Fig.\ref{fig:ps}) to provide the description of temporal resources for a particular partition.

The template declarations in \uppaal support the execution of a \uppTemp{Partition}\uppTemp{Scheduler} model. The parameter \uppVar{pid} of \uppTemp{PartitionScheduler} is the identifier of its partition and the partition schedule is recorded in an array of structures \uppConst{PartitionWindows}. Each element in the array contains two integer fields \uppVar{offset} and \uppVar{duration}, where \uppVar{offset} is the start time of a partition window and \uppVar{duration} denotes the duration of this window. By reading \uppConst{Partition}\uppConst{Windows}\uppConst{Table} from the declarations, the functions \uppFunc{winStart} and \uppFunc{winEnd} with the same integer parameter \uppVar{wind} return the start time and the end time of the $wind$th partition window, respectively. The integer constant \uppConst{MajorFrame} stands for the major time frame $T_{mf}$, and the clock \uppClock{x} measures time within every $T_{mf}$. In the template, all the guards and invariants use \uppClock{x} to control the transitions between locations.

There are three locations in a \uppTemp{PartitionScheduler} model. The initial location \uppLoc{Init} represents a conditional control structure that determines the next location at the start of a major time frame. If a partition window and the major time frame start simultaneously, the model will move to the location \uppLoc{InPartition}. Otherwise, it will enter the location \uppLoc{OutOfPartition}. Within a major time frame, the model keeps traveling between \uppLoc{InPartition} and \uppLoc{OutOfPartition} according to whether or not the current time is in a partition window. For any time from the initial instant, if the \uppTemp{PartitionScheduler} model of \uppVar{pid} enters a new partition window, it will move to the location \uppLoc{InPartition}, and notify the unique task scheduler model in \uppVar{pid} through the output channel \uppSync{enter_partition}. On the contrary, if the \uppTemp{PartitionScheduler} leaves its current partition window, it will move to the location \uppLoc{OutOfPartition}, and send notification to the task scheduler model through the output channel \uppSync{exit_partition}.

\begin{wrapfigure}{r}{0.55\textwidth}
\vspace{-20pt}
\centering
\includegraphics[width=3.5in]{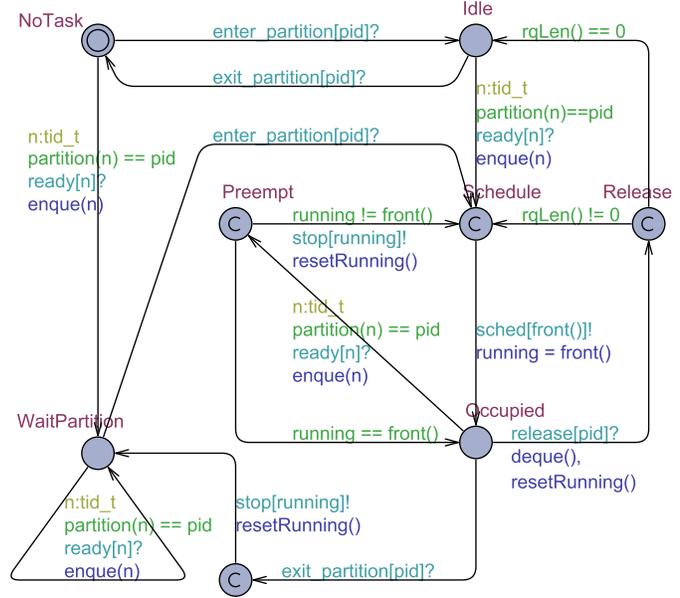}
\caption{TaskScheduler Model}
\label{fig:ts}
\vspace{-10pt}
\end{wrapfigure}

{\small\textbf{TaskScheduler template}} For any partition, there is a task scheduler that executes the preemptive FP scheduling policy while the partition is active. The behavior of the task scheduler is depicted in the TA template \uppTemp{TaskScheduler} (See Fig.\ref{fig:ts}). The only template parameter \uppVar{pid} is the identifier of the task scheduler's partition.

The model of \uppTemp{TaskScheduler} receives notification from the \uppTemp{PartitionScheduler} model through two channels \uppSync{enter_parti}\-\uppSync{tion} and \uppSync{exit_partition}, and uses the channels \uppSync{ready}, \uppSync{release}, \uppSync{sched} and \uppSync{stop} as scheduling commands to manage the tasks in the partition \uppVar{pid}. If there is a task becoming ready to run or relinquishing the processor, the task model will send its \uppTemp{TaskScheduler} model a \uppSync{ready} or \uppSync{release} command respectively. \uppTemp{TaskScheduler} maintains a ready queue that keeps all the tasks ready and waiting to run, and always allocates the processor to the first task with the highest priority in the ready queue. If a new task having a higher priority than any tasks in the ready queue is ready, \uppTemp{TaskScheduler} will insert the task into the ready queue, interrupt the currently running task via the channel \uppSync{stop} and schedule the new selected task via the channel \uppSync{sched}. The task identifier is delivered by the offset of channel arrays in the synchronization between \uppTemp{TaskScheduler} and the task layer.

The ready queue is implemented by the integer array \uppVar{rq} which contains a sorted set of task identifiers in priority order. The tasks with identical priority are served in order of readiness. The function \uppFunc{rqLen} returns the number of the tasks in \uppVar{rq}. We use the function \uppFunc{enque} to insert a new task (identifier) into the ready queue \uppVar{rq} and reorder the tasks in the queue. The function \uppFunc{deque} removes the first element from the ready queue. The first element in \uppVar{rq}, namely the identifier of the currently running task, is returned from the function \uppFunc{front} and recorded in the integer variable \uppVar{running}.

\begin{wraptable}{r}{0.52\textwidth}
\centering
\vspace{-20pt}
\caption{The Major Locations in Task Scheduler}
\label{tab:ts}
\begin{tabular}{l|c|c|c|c}
\hline\hline
\multirow{2}{*}{Location}   &   \multicolumn{2}{c|}{Partition Windows}      &   \multicolumn{2}{c}{Ready Tasks} \\\cline{2-5}
                            &       Outside     &           Inside          &       $\ \ \ 0\ \ \ $         &   $>0$        \\
\hline
{\small\uppLoc{NoTask}}             &       $\surd$     &                           &       $\surd$     &               \\
{\small\uppLoc{Idle}}               &                   &           $\surd$         &       $\surd$     &               \\
{\small\uppLoc{WaitPartition}}      &       $\surd$     &                           &                   &   $\surd$     \\
{\small\uppLoc{Occupied}}           &                   &           $\surd$         &                   &   $\surd$     \\
\hline\hline
\end{tabular}
\vspace{-10pt}
\end{wraptable}

According to whether the current time is in the partition windows as well as to the number of the tasks in the ready queue, we create four major locations listed in Table \ref{tab:ts}. These four locations cover all situations, where the model must be at one of these locations for any time from the initial instant. In contrast, all the other locations of the template are committed and utilized to realize conditional branches or atomic action sequences.

\subsection{Task Layer Models}\label{taskmodel}

We build two SWA templates \uppTemp{PeriodicTask} and \uppTemp{SporadicTask} in \uppaal. Both templates share the same skeleton. So we take \uppTemp{PeriodicTask} for example to sketch out the structure of a task model.

In the template, we define two normal clock \uppClock{x} and \uppClock{curTime} and a stopwatch \uppClock{exeTime}. The clock \uppClock{x} measures the delays prescribed by the task type to calculate the release points of the task. The clock \uppClock{curTime} is used to determine the start of the next task period. By contrast, the stopwatch \uppClock{exeTime} measures the processing time during the execution of an abstract instruction that describes concrete task behavior, and thus progresses only when the model is at the location \uppLoc{Running}.

\begin{figure*}[!t]
\centering
\includegraphics[width=3.5in]{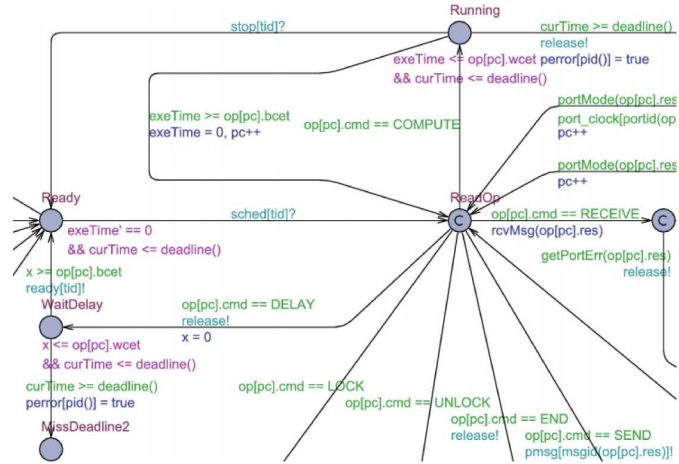}
\caption{The Main Structure of a Task Model}
\label{fig:taskmodel}
\end{figure*}

Once the task is scheduled by \uppTemp{TaskScheduler} through the channel \uppSync{sched}, it will start execution on the processor and move from the location \uppLoc{Ready} to \uppLoc{ReadOp}. For any task in the system, a sequential list of abstract instructions is implemented as the structure array \uppConst{op}. By using an integer variable \uppVar{pc} as a program counter, the task can fetch the next abstract instruction from \uppConst{op[pc]} at the location \uppLoc{ReadOp} (See Fig.\ref{fig:taskmodel}).

According to the command in the abstract instruction currently read from \uppConst{op}, the task model performs a conditional branch and moves from the location \uppLoc{ReadOp} to one of the different locations that represent different operations. Therefore, the command set containing the following seven elements divides the rest of the template into seven corresponding parts.

\begin{itemize}
  \item {\small\textbf{COMPUTE Command:}} When the model reads a \uppConst{COMPUTE} command, it will (re)start the stopwatch \uppClock{exeTime} and enter the location \uppLoc{Running}, which means that the processor is being occupied by the task and executing a computation instruction.
  \item {\small\textbf{LOCK Command:}} By reading a \uppConst{LOCK} command, the task model attempts to acquire the mutual exclusion lock that is specified by the \uppVar{res} field of the instruction. The availability of a mutual exclusion lock depend on the priority ceiling protocol.
  \item {\small\textbf{UNLOCK Command:}} When fetching an \uppConst{UNLOCK} command from \uppConst{op}, the task releases the lock in the instruction and wakes up all the tasks blocked on this lock.
  \item {\small\textbf{DELAY Command:}} The instruction with a \uppConst{DELAY} command can make a task suspended at the location \uppLoc{WaitDelay} for a specified period of time.
  \item {\small\textbf{SEND and RECEIVE Command:}} The commands \uppConst{SEND} and \uppConst{RECEIVE} represent non-blocking message I/O operations among different partitions.
  \item {\small\textbf{END Command:}} The command \uppConst{END} denotes the accomplishment of the current job in this task period. The task will relinquish the processor through the channel \uppSync{release} and stay at the location \uppLoc{WaitNextRelease} until the next period starts.
\end{itemize}

\subsection{Communication Layer Models}

The communication layer consists of four templates: \uppTemp{IPTx} and \uppTemp{IPRx} calculate the delivery latency of the UDP/IP layer. \uppTemp{VLinkTx} and \uppTemp{VLinkRx} calculate the transit delay of frames through a specified VL. We take \uppTemp{VLinkTx} for example. It calculates the latency of frame delivery through the source ES.

\uppTemp{VLinkTx} has a template parameter \uppVar{vlid} that is the unique identifier of a VL. The VL models read the configuration from the array \uppConst{vlink}, which contains a source FIFO buffer \uppVar{src}, an array \uppVar{dst} of destination FIFO buffers, an identifier \uppVar{es} of the VL's source ES, an integer field \uppVar{BAG} that stands for the Bandwidth Allocation Gap (BAG)\cite{arinc664p7}, and an integer field \uppVar{TxDelay} denoting the frame delay\cite{arinc664p7}.

The total delay through a VL is divided into technological latency and configuration latency. The technological latency is independent of traffic load, whereas the configuration latency depends on system configuration and traffic load.

We declare two integer constants \uppConst{TechMin} and \uppConst{TechMax} to be the interval of the technological latency $[\uppConst{TechMin},\uppConst{TechMax}]$. The configuration latency is divided into three parts: the fixed frame delay, the floating delay in waiting for the interval of BAG, and the varying configuration jitter within each BAG. According to the ARINC-664 Part 7\cite{arinc664p7}, a VL should regulate its traffic to send no more than one frame in each BAG. A clock \uppClock{t} measures the interval of the jitter as well as the BAG. By contrast, the configuration jitter within BAGs is caused by the interference from the frames of the other VLs in the same transmitting ES\cite{arinc664p7}. We define an integer array \uppVar{txjitter} where each element provides the maximum configuration jitter according to the current traffic at the output of an ES.

\begin{figure*}[!t]
\centering
\includegraphics[width=4.8in]{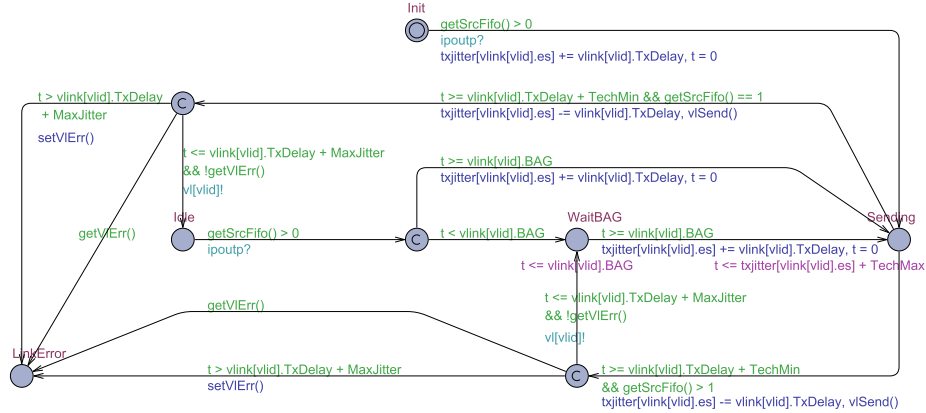}
\caption{The VLinkTx Model}
\label{fig:vltx}
\end{figure*}

As is depicted in Fig.\ref{fig:vltx}, \uppTemp{VLinkTx} obtains notification of packet-receiving on the input channel \uppSync{ipoutp}. At the initial location \uppLoc{Init}, \uppTemp{VLinkTx} waits for the first packet to arrive at the source FIFO. On receiving this first packet, the model enters the location \uppLoc{Sending} and resets the clock \uppClock{t} to start the latency calculation as well as a new BAG interval. Leaving the location \uppLoc{Sending} means the model completes the sending operation of a frame. At this point, \uppTemp{VLinkTx} must invoke the function \uppFunc{vlSend} to decrease the message counter of the source FIFO.

According to the number of packets in the source FIFO, \uppTemp{VLinkTx} waits for the next BAG interval or the next incoming packet after completing a sending operation. First, if the model still has at least one packet in the source FIFO to transmit, it will enter the location \uppLoc{WaitBAG}, thereby waiting for the start of the next BAG. Second, if there is only the sent packet in the source FIFO, the model will stay at the location \uppLoc{Idle} until the arrival of the next incoming packet.

\section{Case Study}\label{sec:experiment}

This section demonstrates the schedulability analysis of an avionics system which combines the workload of \cite{carnevali2013compositional} and the AFDX configuration of \cite{gutierrez2014holistic}. The workload is comprised of 5 partitions, and further divided into 18 periodic tasks and 4 sporadic tasks. Considering the inter-partition messages in the workload, we assign each message type $\mathit{Msg_i},i=\{1,2,3,4\}$ a separate VL with the same subscript. The messages of $\mathit{Msg_1}$ and $\mathit{Msg_2}$ are handled at the refresh period $50ms$ in sampling ports. $\mathit{Msg_3}$ and $\mathit{Msg_4}$ are configured to operate in queuing ports, each of which can accommodate a maximum of one message.

\begin{figure*}[!h]
\begin{minipage}{.5\textwidth}
\centering
\includegraphics[width=2.9in]{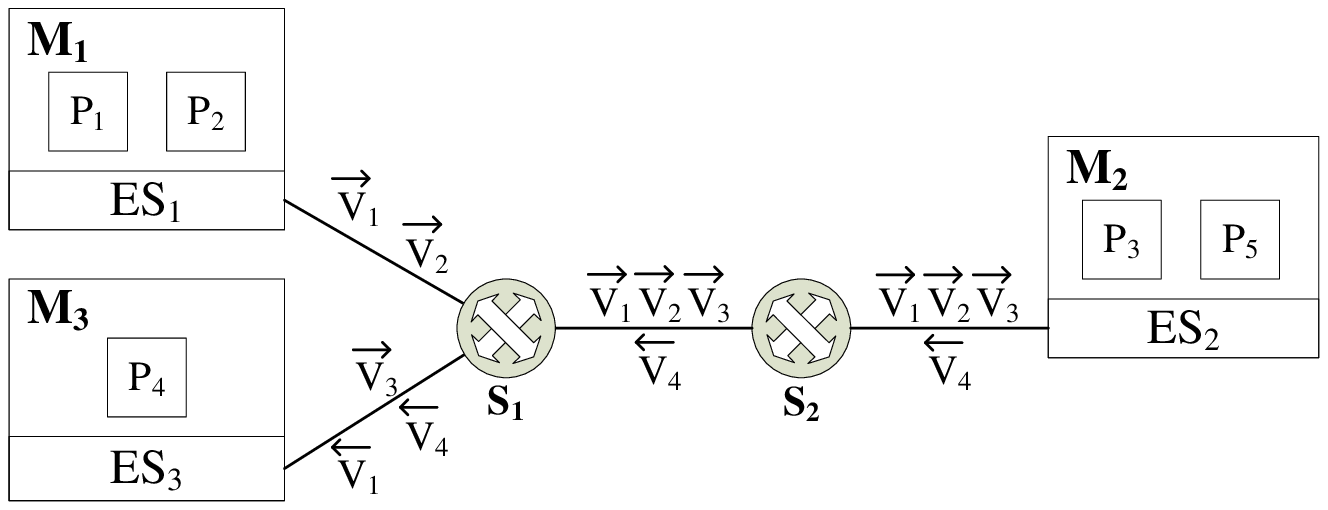}
\end{minipage}%
\begin{minipage}{.5\textwidth}
\centering
\includegraphics[width=2.5in]{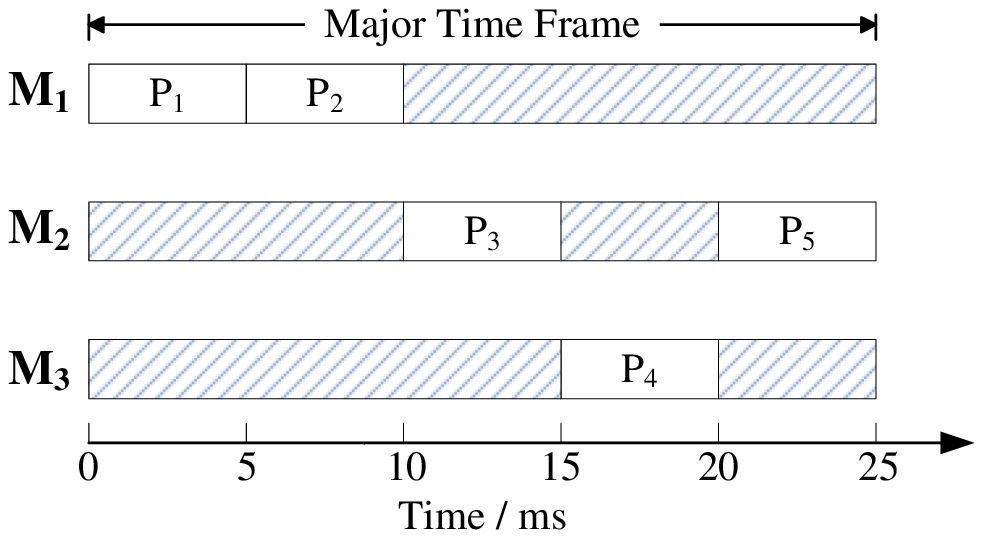}
\end{minipage}%
\caption{The Distributed Avionics Deployment and Partition Schedules (Times in Milliseconds)}
\label{fig:environ}
\label{fig:partsched}
\end{figure*}

Fig.\ref{fig:environ} illustrates the distributed deployment of the workload. We consider 3 ARINC-653 modules connected by an AFDX network, and allocate each partition to one of the modules. The module $M_1$ accommodates $P_1$ and $P_2$, the module $M_2$ executes $P_3$ and $P_5$, and the partition $P_4$ is allocated to $M_3$. There are 4 VLs $V_1$-$V_4$ connecting 3 ESs across 2 switches $S_1$ and $S_2$ in the AFDX network. The arrows above VLs' names indicate the direction of message flow.

The avionics system equips each of its processor cores with a partition schedule. Assume the modules in the experiment to be single-processor platforms. Fig.\ref{fig:partsched} gives the partition schedules, which fix a common major time frame $T_{mf}$ at $25ms$ and allocate $5ms$ to each partition within every $T_{mf}$. All the partition schedules are enabled at the same initial instant and their clocks are always synchronized. The scheduling configuration keeps the temporal order of the partitions in \cite{carnevali2013compositional}. Hence the partition schedules contain five disjoint windows $\langle P_1,0,5\rangle$, $\langle P_2,5,5\rangle$, $\langle P_3,10,5\rangle$, $\langle P_4,15,5\rangle$, and $\langle P_5,20,5\rangle$, where the second parameter is the offset from the start of $T_{mf}$ and last the duration.

After combining all the models of the system, we executed the schedulability analysis in \uppaal. We set $M=100000$ and $\theta=0.001$ for Eq.(\ref{eq:schqry}). The experiment was performed on the \uppaal 4.1.19 64-bit version and an Intel Core i7-5600U laptop processor.

\vspace{-10pt}
\subsubsection*{Results of the Analysis}

The result (The case 1 in Table \ref{tab:timemem}) shows that the above scheduling configuration fails the SMC test and thus is non-schedulable. We can explore the cause of non-schedulability on the basis of counter-examples to help refine the system configuration.

\begin{figure*}
\centering
\includegraphics[width=4.8in]{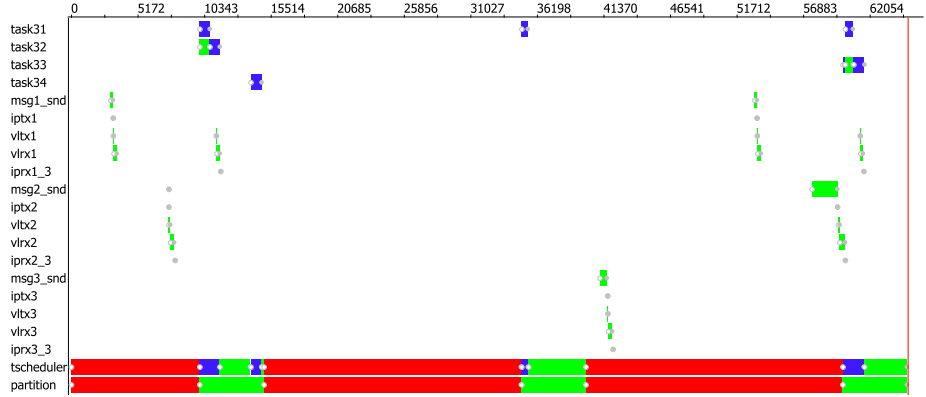}
\caption{The Gantt Chart of a Counter-example (Times in Microseconds)}
\label{fig:err}
\end{figure*}

The Gantt chart in Fig.\ref{fig:err} shows such a counter-example, where the task $\mathit{Tsk^3_2}$ in $P_3$ violates the constraint of the refresh period of $\mathit{Msg_2}$. At the top of the chart are task models, where the line is painted in green whenever a task stays at \uppLoc{Ready} state and in blue at \uppLoc{Running}. The bottom line labels ``partition'' and ``tscheduler'' represent two scheduling-layer models \uppTemp{PartitionScheduler} and \uppTemp{Task}\uppTemp{Scheduler} respectively. For the line ``partition'', color red denotes the time outside $P_3$, and green is within $P_3$. The communication-layer models transmitting the messages of $Msg_\emph{k}$ correspond to the chart lines ``msg\emph{k}\_snd'', ``iptx\emph{k}'', ``iprx\emph{k}'', ``vltx\emph{k}'', and ``vlrx\emph{k}'', which denote the message-delivery delays of $Msg_\emph{k}$.

The counter-example illustrates that network latency increases the risk of breaching the schedulability constraints. Let $t$ be the elapsed time since the initial instant $t_0=0$ shown in the Gantt chart. The first message of $\mathit{Msg_2}$ was sent by the message interface \uppVar{msg2_snd} at $t=7.625$ms, and reached the destination port at $t=8.088$ms. When $\mathit{Tsk^3_2}$ was scheduled to read $\mathit{Msg_2}$ at $t=60.000$ms, the age of the first received message indicated the value $51.912$ms that had exceeded the refresh period. Thus, the copied message of $\mathit{Msg_2}$ was not a valid data sample. Although \uppVar{msg2_snd} sent a new $\mathit{Msg_2}$ message at $t=59.585$ms, the message did not arrive at the destination port until $t=60.184$ms due to network latency.

Considering the effect of network latency on the scheduling configuration, we updated the partition schedules by performing a swap of time slots between $P_1$ and $P_2$. The modified partition schedules provide five windows $\langle P_1,5,5\rangle$, $\langle P_2,0,5\rangle$, $\langle P_3,10,5\rangle$, $\langle P_4,15,5\rangle$, and $\langle P_5,20,5\rangle$. The schedulability analysis of the updated system was executed again. The result (The case 2 in Table \ref{tab:timemem}) shows that the configuration goes through the global SMC test and compositional verification of classical MC. Thus, the updated system finally achieves schedulability.

Table \ref{tab:timemem} shows the execution time and memory usage. In compositional analysis (MC in Table \ref{tab:timemem}), the partition $P_3$ contains more instantiated models (19 processes) than the other four partitions. As a result, model-checking runs slower and requires more memory than the others. Nevertheless, the compositional analysis could be performed on ordinary computers within an acceptable time.

Compared with the compositional way, global analysis based on the same \uppaal models would require 51 processes including all the 22 task models whose state space is much more complex than the others. This causes \uppaal classic to run out of memory within a few minutes, and thus makes the global analysis using classical MC infeasible. In contrast, SMC testing can be quickly accomplished when we perform global analysis (SMC in Table \ref{tab:timemem}), offering effective state space reduction.

\section{Conclusion}\label{sec:conclusion}

In this paper, we present a modeling framework for schedulability analysis of DIMA systems, which are implemented as a set of stopwatch automata in \uppaal. We analyze the \uppaal models including computation and communication by both classical and statistical model checking. The techniques presented in this paper are applicable to the design of DIMA scheduling systems. The experimental results show the applicability of our modeling framework. As future work, we plan to develop a model-based approach to the automatic optimization and generation of a DIMA scheduling system.
\begin{table}[!t]
\centering
\caption{The Experiment Results (Result), Execution Time (Time/sec.) and Memory Usage (Mem/MB)}
\label{tab:timemem}
\begin{tabular}{c|c|r|r|c|c|c|c|c|r|r|c|c|c}
\hline\hline
\multicolumn{7}{c|}{Case 1}                            &\multicolumn{7}{c}{Case 2} \\
\hline
\multicolumn{4}{c|}{\small MC}&\multicolumn{3}{c|}{\small SMC} &\multicolumn{4}{c|}{\small MC}&\multicolumn{3}{c}{\small SMC}\\
\hline
{\footnotesize No.}&{\footnotesize Result}&\multicolumn{1}{c|}{{\footnotesize Time}}&\multicolumn{1}{c|}{{\footnotesize Mem}}           &{\footnotesize Result}&\multicolumn{1}{c|}{{\footnotesize Time}}&\multicolumn{1}{c|}{{\footnotesize Mem}}         &{\footnotesize No.}&{\footnotesize Result}&\multicolumn{1}{c|}{{\footnotesize Time}}&\multicolumn{1}{c|}{{\footnotesize Mem}}           &{\footnotesize Result}&\multicolumn{1}{c|}{{\footnotesize Time}}&\multicolumn{1}{c}{{\footnotesize Mem}}   \\
\hline
{\small $P_1$}       &{\small Yes}&{\small 7.35}&{\small 141}&\multirow{5}{*}{{\small No}}&\multirow{5}{*}{{\small 2.67}}&\multirow{5}{*}{{\small 53}}   &{\small $P_1$}&{\small Yes}&{\small 6.07}&{\small 101}&\multirow{5}{*}{{\small Yes}}&\multirow{5}{*}{{\small 77.58}}&\multirow{5}{*}{{\small 53}}            \\
{\small $P_2$}                   &{\small Yes}&{\small 1.02}&{\small 45}&&&          &{\small $P_2$}&{\small Yes}&{\small 1.09}&{\small 49}&&&             \\
{\small $P_3$}                   &{\footnotesize Maynot}&{\small 57.84}&{\small 563}&&&    &{\small $P_3$}&{\small Yes}&{\small 437.99}&{\small 3150}&&&         \\
{\small $P_4$}                   &{\small Yes}&{\small 0.83}&{\small 45}&&&          &{\small $P_4$}&{\small Yes}&{\small 0.88}&{\small 43}&&&             \\
{\small $P_5$}                   &{\small Yes}&{\small 33.27}&{\small 526}&&&        &{\small $P_5$}&{\small Yes}&{\small 179.25}&{\small 2078}&&&         \\
\hline\hline
\end{tabular}
\end{table}

\bibliographystyle{eptcs}
\bibliography{dima}

\begin{thebibliography}{10}
\providecommand{\bibitemdeclare}[2]{}
\providecommand{\surnamestart}{}
\providecommand{\surnameend}{}
\providecommand{\urlprefix}{Available at }
\providecommand{\url}[1]{\texttt{#1}}
\providecommand{\href}[2]{\texttt{#2}}
\providecommand{\urlalt}[2]{\href{#1}{#2}}
\providecommand{\doi}[1]{doi:\urlalt{http://dx.doi.org/#1}{#1}}
\providecommand{\bibinfo}[2]{#2}

\bibitemdeclare{techreport}{arinc664p7}
\bibitem{arinc664p7}
\bibinfo{author}{\surnamestart AEEC\surnameend} (\bibinfo{year}{2009}):
  \emph{\bibinfo{title}{Aircraft Data Network, Part 7, Avionics Full-Duplex
  Switched Ethernet Network}}.
\newblock \bibinfo{type}{{ARINC} Specification 664P7-1},
  \bibinfo{institution}{Aeronautical Radio Inc.}

\bibitemdeclare{techreport}{arinc653}
\bibitem{arinc653}
\bibinfo{author}{\surnamestart AEEC\surnameend} (\bibinfo{year}{2010}):
  \emph{\bibinfo{title}{Avionics Application Software Standard Interface: Part
  1 - Required Services}}.
\newblock \bibinfo{type}{{ARINC} Specification 653P1-3},
  \bibinfo{institution}{Aeronautical Radio Inc.}

\bibitemdeclare{inproceedings}{amnell2003times}
\bibitem{amnell2003times}
\bibinfo{author}{Tobias \surnamestart Amnell\surnameend},
  \bibinfo{author}{Elena \surnamestart Fersman\surnameend},
  \bibinfo{author}{Leonid \surnamestart Mokrushin\surnameend},
  \bibinfo{author}{Paul \surnamestart Pettersson\surnameend} \&
  \bibinfo{author}{Wang \surnamestart Yi\surnameend}:
  \emph{\bibinfo{title}{TIMES: a tool for schedulability analysis and code
  generation of real-time systems}}.
\newblock In: {\sl \bibinfo{booktitle}{FORMATS 2003}},
  \doi{10.1007/978-3-540-40903-8_6}.

\bibitemdeclare{book}{annighofer2014systems}
\bibitem{annighofer2014systems}
\bibinfo{author}{Bj{\"o}rn \surnamestart Annigh{\"o}fer\surnameend} \&
  \bibinfo{author}{Frank \surnamestart Thielecke\surnameend}
  (\bibinfo{year}{2014}): \emph{\bibinfo{title}{A Systems Architecting
  Framework for Distributed Integrated Modular Avionics}}.
\newblock \bibinfo{publisher}{DGLR}, \doi{10.1007/s13272-015-0156-1}.

\bibitemdeclare{inproceedings}{boudjadar2014compositional}
\bibitem{boudjadar2014compositional}
\bibinfo{author}{Jalil \surnamestart Boudjadar\surnameend},
  \bibinfo{author}{Kim~Guldstrand \surnamestart Larsen\surnameend},
  \bibinfo{author}{Jin~Hyun \surnamestart Kim\surnameend} \&
  \bibinfo{author}{Ulrik \surnamestart Nyman\surnameend}:
  \emph{\bibinfo{title}{Compositional schedulability analysis of an avionics
  system using {UPPAAL}}}.
\newblock In: {\sl \bibinfo{booktitle}{AASE 2014}}.

\bibitemdeclare{inproceedings}{carnevali2011formal}
\bibitem{carnevali2011formal}
\bibinfo{author}{Laura \surnamestart Carnevali\surnameend},
  \bibinfo{author}{Giuseppe \surnamestart Lipari\surnameend},
  \bibinfo{author}{Alessandro \surnamestart Pinzuti\surnameend} \&
  \bibinfo{author}{Enrico \surnamestart Vicario\surnameend}:
  \emph{\bibinfo{title}{A formal approach to design and verification of
  two-level Hierarchical Scheduling systems}}.
\newblock In: {\sl \bibinfo{booktitle}{RST 2011}}, \doi{10.1007/BF00360340}.

\bibitemdeclare{article}{carnevali2013compositional}
\bibitem{carnevali2013compositional}
\bibinfo{author}{Laura \surnamestart Carnevali\surnameend},
  \bibinfo{author}{Alessandro \surnamestart Pinzuti\surnameend} \&
  \bibinfo{author}{Enrico \surnamestart Vicario\surnameend}
  (\bibinfo{year}{2013}): \emph{\bibinfo{title}{Compositional verification for
  hierarchical scheduling of real-time systems}}.
\newblock {\sl \bibinfo{journal}{IEEE Transactions on Software Engineering}}
  \bibinfo{volume}{39}(\bibinfo{number}{5}), pp. \bibinfo{pages}{638--657},
  \doi{10.1109/TSE.2012.54}.

\bibitemdeclare{inproceedings}{cassez2000impressive}
\bibitem{cassez2000impressive}
\bibinfo{author}{Franck \surnamestart Cassez\surnameend} \&
  \bibinfo{author}{Kim \surnamestart Larsen\surnameend}:
  \emph{\bibinfo{title}{The impressive power of stopwatches}}.
\newblock In: {\sl \bibinfo{booktitle}{CONCUR 2000}},
  \doi{10.1007/3-540-44618-4_12}.

\bibitemdeclare{inproceedings}{cicirelli2012development}
\bibitem{cicirelli2012development}
\bibinfo{author}{Franco \surnamestart Cicirelli\surnameend},
  \bibinfo{author}{Angelo \surnamestart Furfaro\surnameend} \&
  \bibinfo{author}{Libero \surnamestart Nigro~et al.\surnameend}:
  \emph{\bibinfo{title}{Development of a schedulability analysis framework
  based on pTPN and {U}PPAAL with stopwatches}}.
\newblock In: {\sl \bibinfo{booktitle}{DSRA 2012}},
  \doi{10.1109/DS-RT.2012.16}.

\bibitemdeclare{article}{david2015uppaal}
\bibitem{david2015uppaal}
\bibinfo{author}{Alexandre \surnamestart David\surnameend},
  \bibinfo{author}{Kim~G \surnamestart Larsen\surnameend},
  \bibinfo{author}{Axel \surnamestart Legay\surnameend},
  \bibinfo{author}{Marius \surnamestart Miku{\v{c}}ionis\surnameend} \&
  \bibinfo{author}{Danny~B{\o}gsted \surnamestart Poulsen\surnameend}
  (\bibinfo{year}{2015}): \emph{\bibinfo{title}{Uppaal SMC tutorial}}.
\newblock {\sl \bibinfo{journal}{STTT}}
  \bibinfo{volume}{17}(\bibinfo{number}{4}), pp. \bibinfo{pages}{397--415},
  \doi{10.1007/s10009-014-0361-y}.

\bibitemdeclare{techreport}{dodd2006coloured}
\bibitem{dodd2006coloured}
\bibinfo{author}{RB~\surnamestart Dodd\surnameend} (\bibinfo{year}{2006}):
  \emph{\bibinfo{title}{Coloured petri net modelling of a generic avionics
  mission computer}}.
\newblock \bibinfo{type}{Technical Report}, \bibinfo{institution}{DTIC}.

\bibitemdeclare{inproceedings}{easwaran2009compositional}
\bibitem{easwaran2009compositional}
\bibinfo{author}{Arvind \surnamestart Easwaran\surnameend},
  \bibinfo{author}{Insup \surnamestart Lee\surnameend}, \bibinfo{author}{Oleg
  \surnamestart Sokolsky\surnameend} \& \bibinfo{author}{Steve \surnamestart
  Vestal\surnameend}: \emph{\bibinfo{title}{A compositional scheduling
  framework for digital avionics systems}}.
\newblock In: {\sl \bibinfo{booktitle}{RTCSA 2009}},
  \doi{10.1109/RTCSA.2009.46}.

\bibitemdeclare{article}{gutierrez2014holistic}
\bibitem{gutierrez2014holistic}
\bibinfo{author}{J~Javier \surnamestart Guti{\'e}rrez\surnameend},
  \bibinfo{author}{J~Carlos \surnamestart Palencia\surnameend} \&
  \bibinfo{author}{Michael~Gonz{\'a}lez \surnamestart Harbour\surnameend}
  (\bibinfo{year}{2014}): \emph{\bibinfo{title}{Holistic schedulability
  analysis for multipacket messages in {AFDX} networks}}.
\newblock {\sl \bibinfo{journal}{Real-Time Systems}}
  \bibinfo{volume}{50}(\bibinfo{number}{2}), \doi{10.1007/s11241-013-9192-2}.

\bibitemdeclare{inproceedings}{sun2014component}
\bibitem{sun2014component}
\bibinfo{author}{Youcheng \surnamestart Sun\surnameend},
  \bibinfo{author}{Giuseppe \surnamestart Lipari\surnameend},
  \bibinfo{author}{Romain \surnamestart Soulat\surnameend},
  \bibinfo{author}{Laurent \surnamestart Fribourg\surnameend} \&
  \bibinfo{author}{Nicolas \surnamestart Markey\surnameend}:
  \emph{\bibinfo{title}{Component-based analysis of hierarchical scheduling
  using linear hybrid automata}}.
\newblock In: {\sl \bibinfo{booktitle}{ERCSA 2014}},
  \doi{10.1109/RTCSA.2014.6910502}.

\bibitemdeclare{inproceedings}{wang2013research}
\bibitem{wang2013research}
\bibinfo{author}{Guoqing \surnamestart Wang\surnameend} \&
  \bibinfo{author}{Qingfan \surnamestart Gu\surnameend}:
  \emph{\bibinfo{title}{Research on distributed integrated modular avionics
  system architecture design and implementation}}.
\newblock In: {\sl \bibinfo{booktitle}{DASC 2013}},
  \doi{10.1109/dasc.2013.6712647}.

\end{thebibliography}

\newpage
\appendix
\section*{Appendix}

The appendix consists of three sections. Appendix A gives a description of the remaining \uppaal models in this paper. Appendix B details the avionics workload in the case study. The AFDX configuration is then presented in Appendix C.

\section{Remaining Models}

\subsection{UDP/IP Layer Models}

Although the behavior of the UDP/IP layer largely depends on the implementation of the network protocol stack, we create two TA templates \uppTemp{IPTx} and \uppTemp{IPRx} to estimate the latency of message delivery through the UDP/IP layer. Both templates have two \uppType{msgbuf_t} parameters \uppVar{src} and \uppVar{dst} that denote the source buffer and the destination buffer respectively. After being instantiated in the system declarations, these two templates give rise to a set of UDP/IP layer models. By operating the message counters \uppVar{buf} in \uppVar{src} and \uppVar{dst}, the models transfer messages from their respective source buffers to the destination buffers.

Two types of message buffers are provided for the UDP/IP layer models. The first is the port buffers between tasks and the UDP/IP layer. The set of port buffers in the system is defined as a global array \uppVar{portbuf}. We declare the second as another global \uppType{msgbuf_t} array \uppVar{fifo} which represents the FIFO buffers between the UDP/IP layer and virtual links. Obviously, \uppTemp{IPTx} and \uppTemp{IPRx} forward messages in mutually opposed directions.

{\small\textbf{The template IPTx}} calculates the latency of message delivery from a port buffer \uppVar{src} to a FIFO buffer \uppVar{dst}. In order to specify the time interval of forwarding a single message from the source port to the destination FIFO, we declare two integer constants \uppConst{IpFwdMin} and \uppConst{IpFwdMax} as the lower bound and upper bound, respectively. For any queuing port, the model can perform IP fragmentation according to the integer parameter \uppVar{frag}, so breaking one message into $frag$ IP fragments during transmission. In this case, If \uppTemp{IPTx} handles a message sent from \uppVar{src} to \uppVar{dst}, the message counter \uppVar{buf} of \uppVar{src} will be decreased by one, and meanwhile the counter of \uppVar{dst} must increase by \emph{frag}. On the contrary, since sampling ports should not use IP fragmentation\cite{arinc664p7}, the default value 1 is assigned to the template parameters \uppVar{frag} of the \uppTemp{IPTx} models whose source ports are declared to be sampling mode.

\begin{figure*}[!h]
\centering
\includegraphics[width=3in]{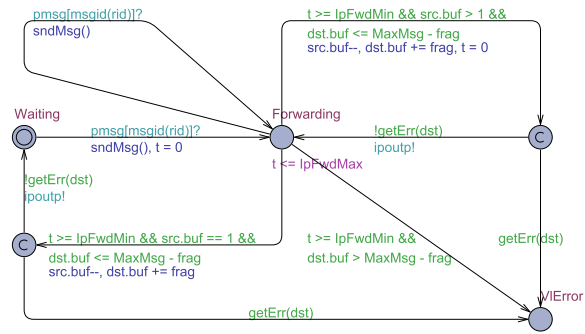}
\caption{The IPTx Model}
\label{fig:iptx}
\end{figure*}

As is depicted in Fig.\ref{fig:iptx}, an iteration structure realizes the major function of \uppTemp{IPTx}. After starting from the initial location \uppLoc{Waiting}, the model keeps waiting for the first incoming message to arrive in the port buffer \uppVar{src}. Once any tasks send the port a message through the channel \uppSync{pmsg}, \uppTemp{IPTx} will increase the counter of \uppVar{src} by invoking the function \uppFunc{sndMsg} and thereupon move to the location \uppLoc{Forwarding}. This location represents that the UDP/IP layer is executing the forwarding operation. Thus, the model can non-deterministically choose a forwarding delay $T_F$ from the interval $[\uppConst{IpFwdMin},\uppConst{IpFwdMax}]$ and stay at the location \uppLoc{Forwarding} for $T_F$ to forward the first message in the \uppVar{src} port. Receiving a new incoming message during forwarding will not affect the delay $T_F$ but only raise the counter of \uppVar{src} once more. When the model completes the current forwarding operation, it will send notification to a VL model through the output channel \uppOut{ipoutp} as well as operate the message counters of \uppVar{src} and \uppVar{dst}. The forwarding operation continues until the source buffer is empty. In other words, if there is still at least one message in the \uppVar{src} port, the model will immediately return to the location \uppLoc{Forwarding} to forward the next message. Otherwise, the model will move back to the initial location \uppLoc{Waiting} and restart the iteration to wait for the arrival of the following messages.

The template \uppTemp{IPTx} has a location \uppLoc{VlError} that represents the existence of errors in the destination VL. First, if any errors are reported by the VL, the function \uppFunc{getErr} serving as a guard will return $true$ and the model will stop the forwarding iteration at the location \uppLoc{VlError}. Second, a shortage of FIFO space will also lead the model to \uppLoc{VlError}. The FIFO \uppVar{dst}, which has a \uppConst{MaxMsg} capacity, should accommodate at least $frag$ IP packet(s) during every forwarding operation, unless the guard \uppGuard{dst.buf > MaxMsg-frag} holds at the location \uppLoc{Forwarding}.

In addition, once the function \uppFunc{sndMsg} tries sending a message to a full queuing port, the \uppTemp{IPTx} model will report an overflow error to the source task, which will thereupon move to an error location \uppLoc{MsgErr}. In contrast, sampling ports can avoid overflow by overwriting the previous message in the buffer. Hence \uppFunc{sndMsg} assigns 1 directly to the counter of a sampling port.

\begin{figure*}[!t]
\centering
\includegraphics[width=4.5in]{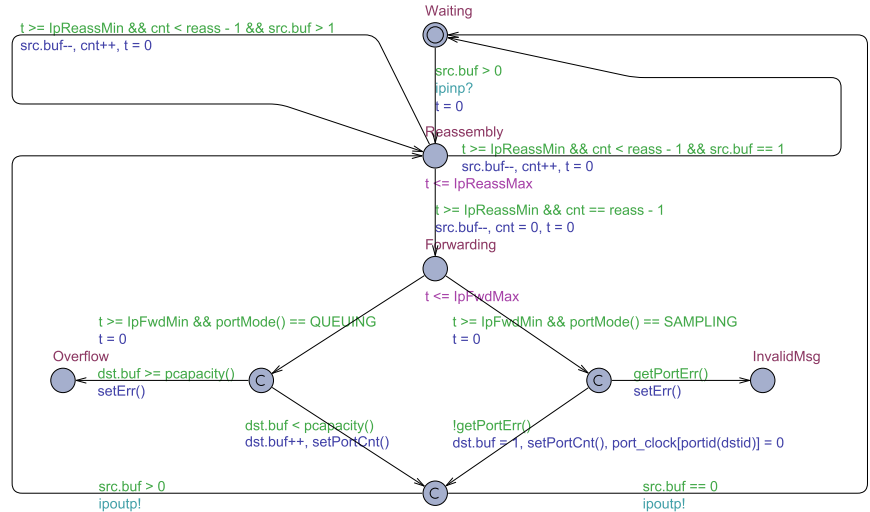}
\caption{The IPRx Model}
\label{fig:iprx}
\end{figure*}

{\small\textbf{The template IPRx}} calculates the latency of message delivery from a FIFO buffer \uppVar{src} to a port buffer \uppVar{dst}. As is shown in Fig.\ref{fig:iprx}, \uppTemp{IPRx} also includes a forwarding iteration similar to \uppTemp{IPTx}, but we insert two following parts into the iteration structure.

The first is a reassembly iteration between the initial location \uppLoc{Waiting} and the location \uppLoc{Forwarding}. The nested iteration contains only one \uppLoc{Reassembly} location, where the model waits for IP packets to arrive and reassembles a complete message. Assume that the IP packets of every message can arrive in order. The template parameter \uppVar{reass} denotes that $reass$ consecutive IP packets constitute one complete message. At the location \uppLoc{Reassembly}, a non-deterministic time between two integer constants \uppConst{IpReassMin} and \uppConst{IpReassMax} is spent in processing an IP fragment. After this processing delay, the IP packet must be removed from the \uppVar{src} FIFO. When reassembling a message, the model uses an integer variable \uppVar{cnt} to record the number of IP packets that have arrived of the message. Once the model accumulates the $reass$ consecutive IP packets, it will enter the location \uppLoc{Forwarding} to forward the complete message to \uppVar{dst}.

Second, according to the transfer mode of the destination port \uppVar{dst}, we add two different paths following the location \uppLoc{Forwarding} to operate queuing ports and sampling ports, respectively. For a queuing port, we increase the value of its message counter \uppVar{buf} by 1. The guard \uppGuard{dst.buf < pcapacity()} ensures that the number of messages in \uppVar{dst} is less than the capacity of \uppVar{dst}. Otherwise, the model will report an overflow error by moving to the \uppLoc{Overflow} location. For a sampling port, we should fill an empty buffer with the new message or overwrite the previous one. Therefore, \uppTemp{IPRx} directly assigns 1 to the message counter \uppVar{buf}. Meanwhile, \uppTemp{IPRx} resets the port clock of \uppVar{dst}. If a task found an invalid message in the \uppVar{dst} after comparing the refresh period of the message with the port clock of the \uppVar{dst}, the \uppTemp{IPRx} model would detect a port error using the function \uppFunc{getPortErr} and stop its forwarding iteration at the location \uppLoc{InvalidMsg}.

Both the templates\uppTemp{IPTx} and \uppTemp{IPRx} provide a typical processing procedure for the UDP/IP protocol. Users can easily adapt the templates for their specific implementation.

\subsection{Virtual Link Models}\label{subsec:vl}

We create two TA templates \uppTemp{VLinkTx} and \uppTemp{VLinkRx} to calculate the transit delay of frames through a specified VL.

\begin{figure*}[!t]
\centering
\includegraphics[width=2in]{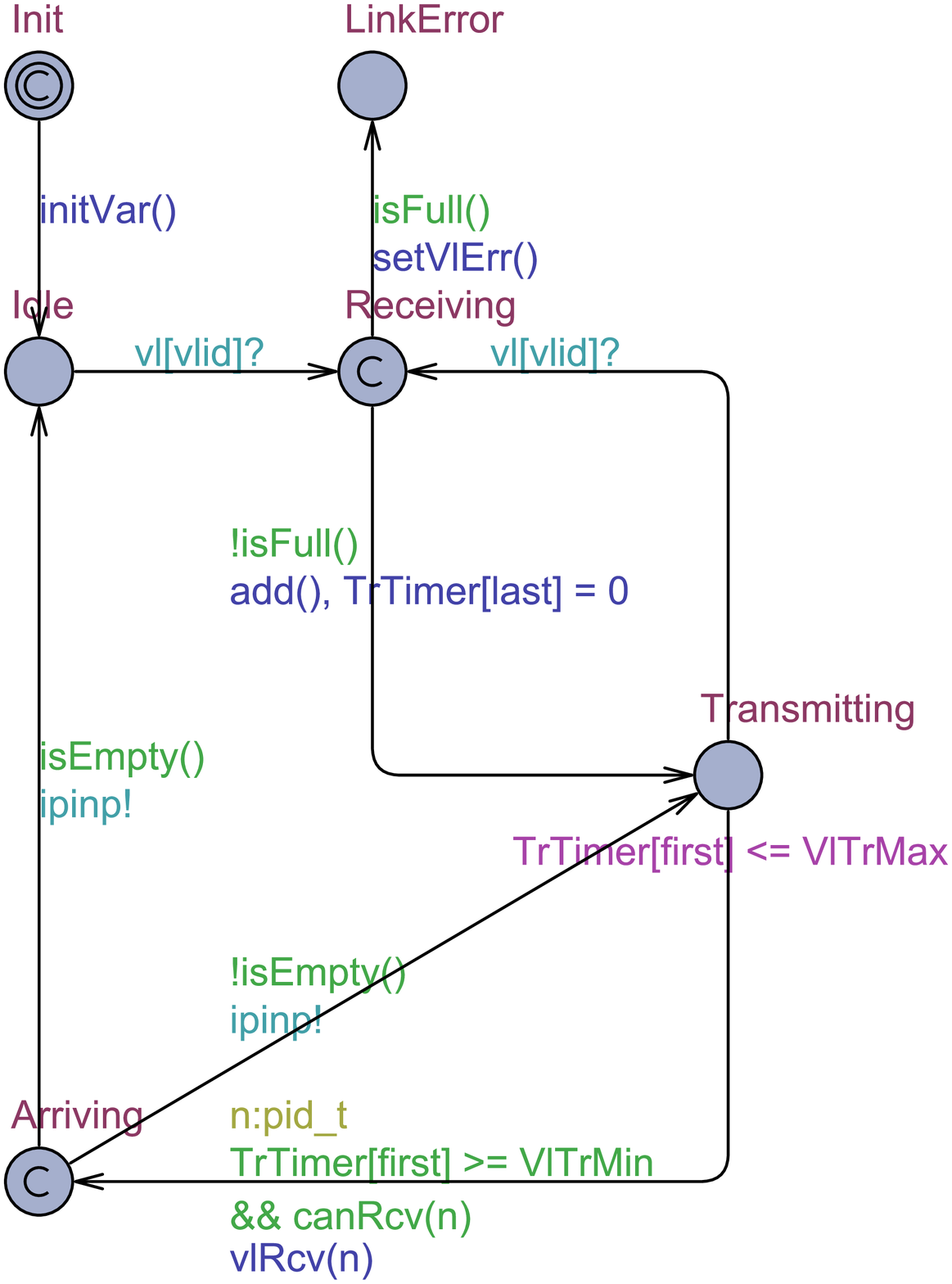}
\caption{The VLinkRx Model}
\label{fig:vlrx}
\end{figure*}

The template \uppTemp{VLinkRx} provides the latency of frame delivery through the route from the first switch to a destination ES. In \uppTemp{VLinkRx}, the integer argument \uppVar{links}(resp. \uppVar{switches}) denotes the number of physical links(resp. switches) along the route. Assume that each switch and ES can send and receive frames at wire speed. We divide the latency into three parts: the transmission delay of physical links, the processing delay of switches, and the latency at the destination ES. First, given the frame delay \uppConst{vlink[vlid].TxDelay} of each physical link, the total transmission delay can be described as the expression $\uppVar{vlink[vlid].TxDelay*links}$. Second, two integer constants \uppConst{SwMin} and \uppConst{SwMax} give the interval $[\uppConst{SwMin}*\uppVar{switches}, \uppConst{SwMax}*\uppVar{switches}]$ of the processing delay through switches. Similarly, we also define two integer constants \uppConst{RxMin} and \uppConst{RxMax} to record the the interval $[\uppConst{RxMin}, \uppConst{RxMax}]$ of the latency at the destination ES. By adding up these three delays, we provide each frame with the range of the total latency through the route [\uppVar{vlink[vlid].TxDelay*links} + \uppVar{SwMin*switches} + \uppConst{RxMin}, \uppVar{vlink[vlid].TxDelay*links} + \uppVar{SwMax*switches} + \uppConst{RxMax}]. At the initial instant, the model should invoke the function \uppFunc{initVar}, which records the total latency interval in two integer variables \uppVar{VlTrMin} and \uppVar{VlTrMax}.

Assume that no more than \uppConst{MaxPackets} frames can travel along the route simultaneously. We create the clock array \uppVar{TrTimer} with a \uppConst{MaxPackets} size to measure the delivery latency of all the frames through the route. Given in-order reliable delivery in the network, \uppVar{TrTimer} is operated as a circular queue, where each valid clock serves as the unique timer of a frame. As long as \uppVar{TrTimer} is not empty, two integer variables \uppVar{first} and \uppVar{last} indicate the start and the end of valid clocks in \uppVar{TrTimer}, respectively. The clock \uppVar{TrTimer[first]} represents the timer of the earliest frame that is travelling along the route. When \uppTemp{VLinkRx} completes the delivery of a frame, \uppVar{TrTimer[first]} will be reset and the offset \uppVar{first} will be also updated according to the circular-queue implementation. By contrast, the clock \uppVar{TrTimer[last]} measures the delivery latency of the most recent frame being transmitted along the route. Immediately a new frame appears in its first physical link, \uppTemp{VLinkRx} will call the function \uppFunc{add} that appends a new element to the array \uppVar{TrTimer} and updates the offset \uppVar{last}.

The template \uppTemp{VLinkRx} is shown in Fig.\ref{fig:vlrx}. After initializing the latency interval $[\uppVar{VlTrMin},\uppVar{VlTrMax}]$ of a single frame in the function \uppFunc{initVar}, the \uppTemp{VLinkRx} model stays at the location \uppLoc{Idle} until it receives frame arrival notification from the input channel \uppIn{vl}. New frame arrivals are handled at the location \uppLoc{Receiving}. First, \uppTemp{VLinkRx} invokes the function \uppFunc{add} to insert a new valid clock in \uppVar{TrTimer[last]}. Immediately afterwards, the model resets the clock \uppVar{TrTimer[last]} as well as enters the location \uppLoc{Transmitting}, thereby starting the latency calculation of the most recent frame. Additionally, if more than \uppConst{MaxPackets} frames were transmitted simultaneously, the model should report error and move to the location \uppLoc{LinkError}.

For each frame that is travelling along the route, \uppTemp{VLinkRx} spends a non-deterministic time between \uppVar{VlTrMin} and \uppVar{VlTrMax} performing the frame delivery at the location \uppLoc{Transmitting}. Since a VL may have more than one destination partitions, we define a select \uppSelect{n:pid_t} to give a random arrival order of multicast frames. During frame delivery, the model still react to the input channel \uppIn{vl} so that we can handle each frame arriving in the VL \uppVar{vlid}. Once a frame is delivered to the UDP/IP layer in the receiving ES, the \uppTemp{VLinkRx} model will call the function \uppFunc{vlRcv} to increase the destination FIFO, and notify \uppVar{vlid}'s \uppTemp{IPRx} model through the output channel \uppOut{ipinp}. Thereafter, frame delivery continues at the location \uppLoc{Transmitting} until there is no frame travelling along the route. In that case, the model will return to the location \uppLoc{Idle} and wait for the incoming frames again.

\subsection{Message Interfaces}\label{sec:mimodel}

\begin{figure*}[!t]
\begin{minipage}[t]{0.5\linewidth}
\centering
\includegraphics[width=2.4in]{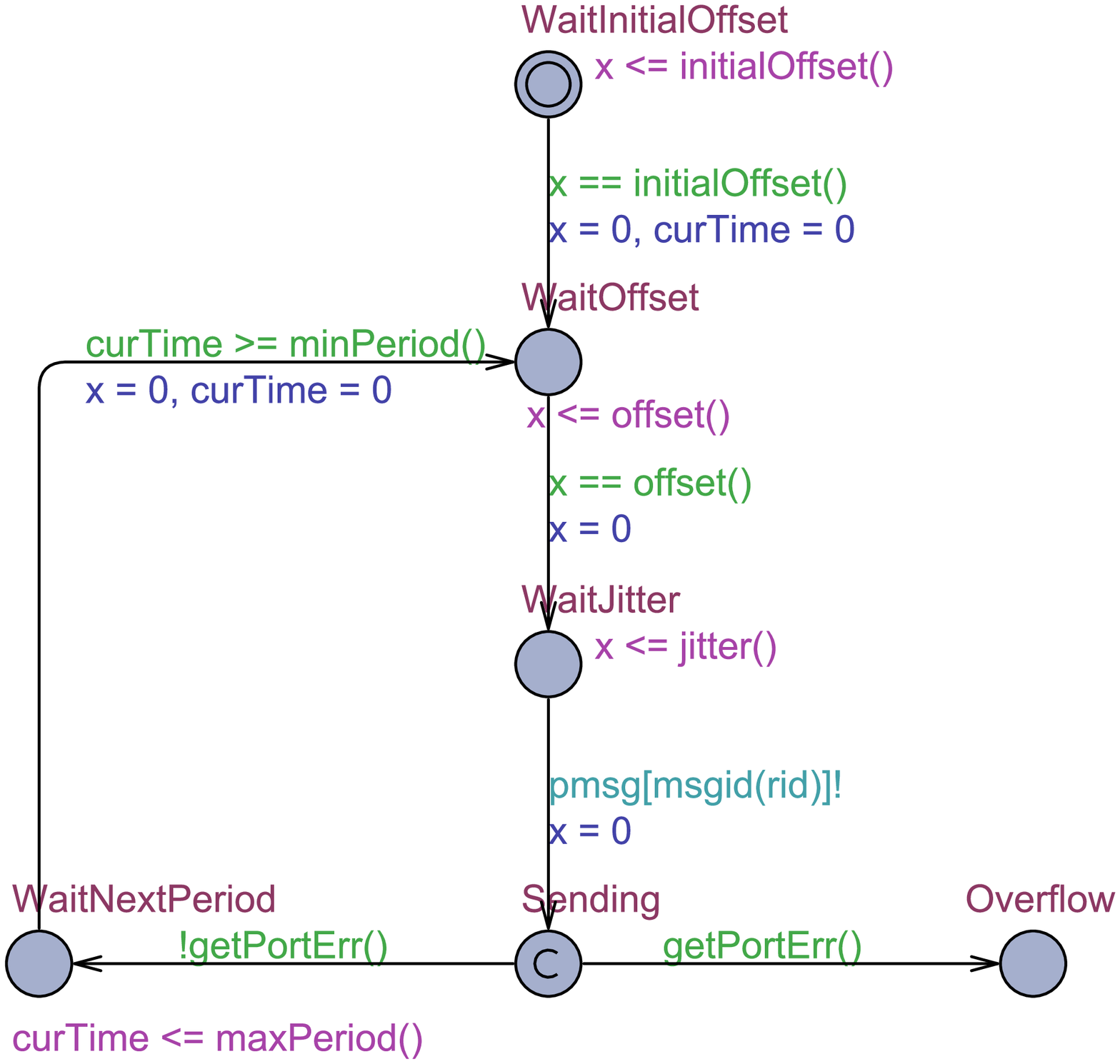}
\caption{Periodic Message Sender}
\label{fig:pmsgsender}
\end{minipage}%
\begin{minipage}[t]{0.5\linewidth}
\centering
\includegraphics[width=2.4in]{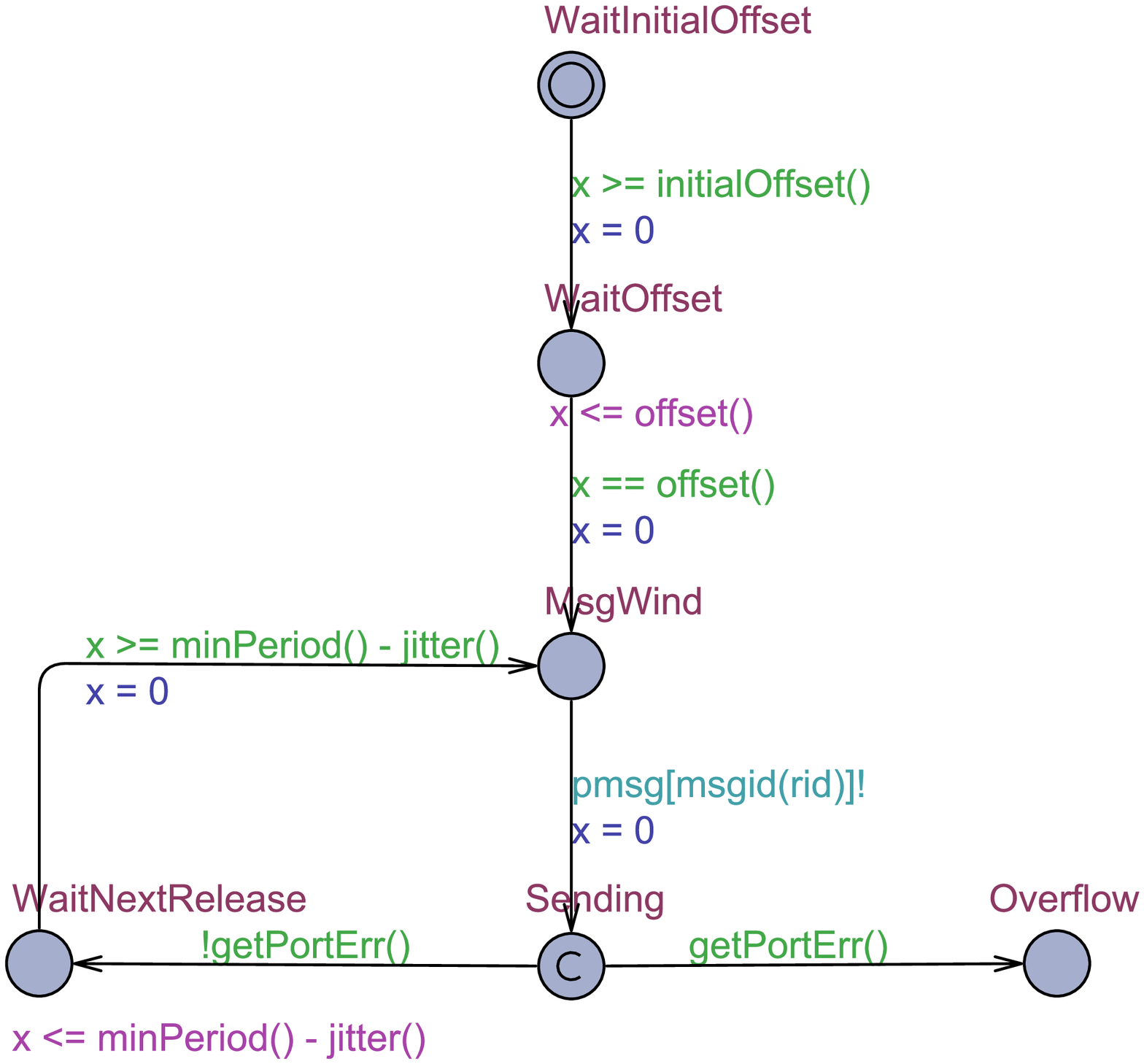}
\caption{Sporadic Message Sender}
\label{fig:smsgsender}
\end{minipage}
\end{figure*}

Two \uppaal TA templates have been created to help the construction of message interfaces. By instantiating the parameters in the templates, one can build a message interface for a particular message type. In general, message-sending actions and the release of their source tasks have similar temporal patterns. Hence we assume that periodic and sporadic tasks generate periodic and sporadic messages respectively.

The TA template of \uppTemp{PMsgSender} depicted in Fig.\ref{fig:pmsgsender} is built for the periodic messages. We define a set of functions in the template to access the parameters of this message pattern. The functions \uppFunc{initialOffset}, \uppFunc{offset} and \uppFunc{jitter} read the corresponding parameters from the declarations. The functions \uppFunc{minPeriod} and \uppFunc{maxPeriod} always return the same value of a fixed period. There are two clocks \uppClock{curTime} and \uppClock{x} in the template. The clock \uppClock{curTime} represents the accumulated time since the beginning of the current period. By using the clock \uppClock{x} to measure time repeatedly in each period, \uppTemp{PMsgSender} can wait for the delays of initial offset, offset and jitter at the locations \uppLoc{WaitInitialOffset}, \uppLoc{WaitOffset} and \uppLoc{WaitJitter} respectively. The template exploits the broadcast action array \uppSync{pmsg} to model the message-sending behavior. When transmitting a message to the UDP/IP layer, periodic message sender model gives the \uppTemp{IPTx} model a notification of the broadcast action \uppSync{pmsg[msgid(rid)]}, where the offset \uppVar{msgid(rid)} is the identifier of the message type. After the broadcast synchronization of \uppSync{pmsg} in normal execution, the model stays at the location \uppLoc{WaitNextPeriod} until the next period starts. However, if the port buffer were overflow after the broadcast synchronization, the model would stop the following message-sending operation and enter the location \uppLoc{Overflow} to indicate the violation of schedulability properties.

As is shown in Fig.\ref{fig:smsgsender}, the other TA template of \uppTemp{SMsgSender} has an analogous structure with the \uppTemp{PMsgSender}. In accordance with the sporadic message pattern, a \uppTemp{SMsgSender} can stay at the location \uppLoc{MsgWind} for any time before its next sending action. Location \uppLoc{WaitNextRelease} enables a minimum separation between the consecutive messages.

\onecolumn
\section{Avionics Workload}

\begin{multicols}{2}
As shown in Table \ref{tab:wl}, the workload is comprised of 5 partitions ($P_1-P_5$), and further divided into 18 periodic tasks and 4 sporadic tasks. The type of a task depends on its \emph{release} interval. A periodic task has a deterministic period, whereas the release time of a sporadic task is only bounded by a minimum separation. The execution of a task is characterized as a sequence of \emph{chunks}. Each chunk involves the description of a non-deterministic \emph{execution time}, required resources and message-passing operations. There are 3 intra-partition locks(column \emph{mutex}) and 4 inter-partition message types defined in the task set. The columns \emph{output} and \emph{input} indicate transfer direction of messages.
\end{multicols}

\begin{table}[!h]
\centering
\caption{The Workload of the Avionics System\cite{carnevali2013compositional,easwaran2009compositional}(Times in Milliseconds)}
\label{tab:wl}
\begin{tabular}{||c||c|c|c|c|c|c||c|c|c|c||}
\hline\hline
\multirow{2}{*}{No.}
&\multirow{2}{*}{Task}      &\multirow{2}{*}{Release}           &\multirow{2}{*}{Offset}    &\multirow{2}{*}{Jitter}    &\multirow{2}{*}{Deadline}      &\multirow{2}{*}{Priority}      &\multicolumn{4}{c||}{Execution Chunks}\\\cline{8-11}&&&&&&&Time  &Mutex  &Output  &Input    \\
\hline\hline
\multirow{7}{*}{$P_1$}
&\multirow{2}{*}{$\mathit{Tsk^1_1}$} &\multirow{2}{*}{[25,25]}           &\multirow{2}{*}{2}         &\multirow{2}{*}{0}         &\multirow{2}{*}{25}            &\multirow{2}{*}{2}             &[0.8,1.3]  &-          &-          &-          \\\cline{8-11}
&&&&&&                          &[0.1,0.2]  &-          &-          &-          \\
\cline{2-11}
&$\mathit{Tsk^1_2}$                  &[50,50]                            &3                          &0                          &50
&3                              &[0.2,0.4]  &-          &$Msg_1$    &-          \\
\cline{2-11}
&$\mathit{Tsk^1_3}$                  &[50,50]                            &3                          &0                          &50
&4                              &[2.7,4.2]  &-          &-          &-          \\
\cline{2-11}
&$\mathit{Tsk^1_4}$                  &[50,50]                            &0                          &0                          &50
&5                              &[0.1,0.2]  &$Mux^1_1$  &-          &-          \\
\cline{2-11}
&\multirow{2}{*}{$\mathit{Tsk^1_5}$} &\multirow{2}{*}{[120,$\infty$)}    &\multirow{2}{*}{0}         &\multirow{2}{*}{0}         &\multirow{2}{*}{120}            &\multirow{2}{*}{6}             &[0.6,0.9]  &-          &-          &-          \\\cline{8-11}
&&&&&&                          &[0.1,0.2]  &$Mux^1_1$  &-          &-          \\
\hline\hline
\multirow{5}{*}{$P_2$}
&$\mathit{Tsk^2_1}$                  &[50,50]                            &0                          &0.5                          &50
&2                              &[1.9,3.0]  &-          &-          &-          \\
\cline{2-11}
&$\mathit{Tsk^2_2}$                  &[50,50]                            &2                          &0                          &50
&3                              &[0.7,1.1]  &-          &$Msg_2$    &-          \\
\cline{2-11}
&$\mathit{Tsk^2_3}$                  &[100,100]                          &0                          &0                          &100
&4                              &[0.1,0.2]  &$Mux^2_1$  &-          &-          \\
\cline{2-11}
&\multirow{2}{*}{$\mathit{Tsk^2_4}$} &\multirow{2}{*}{[100,$\infty$)}    &\multirow{2}{*}{10}         &\multirow{2}{*}{0}         &\multirow{2}{*}{100}            &\multirow{2}{*}{5}             &[0.8,1.3]  &-          &-          &-          \\\cline{8-11}
&&&&&&                          &[0.2,0.3]  &$Mux^2_1$  &-          &-          \\
\hline\hline
\multirow{5}{*}{$P_3$}
&$\mathit{Tsk^3_1}$                  &[25,25]                            &0                          &0.5                          &25
&2                              &[0.5,0.8]  &-          &-          &$Msg_1$    \\
\cline{2-11}
&$\mathit{Tsk^3_2}$                  &[50,50]                            &0                          &0                          &50
&3                              &[0.7,1.1]  &-          &-          &$Msg_2$    \\
\cline{2-11}
&$\mathit{Tsk^3_3}$                  &[50,50]                            &0                          &0                          &50
&4                              &[1.0,1.6]  &-          &-          &$Msg_3$    \\
\cline{2-11}
&\multirow{2}{*}{$\mathit{Tsk^3_4}$} &\multirow{2}{*}{[100,$\infty$)}    &\multirow{2}{*}{11}         &\multirow{2}{*}{0}         &\multirow{2}{*}{100}            &\multirow{2}{*}{5}             &[0.7,1.0]  &-          &-          &-          \\\cline{8-11}
&&&&&&                          &[0.1,0.3]  &-          &-          &-          \\
\hline\hline
\multirow{5}{*}{$P_4$}
&$\mathit{Tsk^4_1}$                  &[25,25]                            &3                          &0.2                          &25
&2                              &[0.7,1.2]  &-          &-          &-          \\
\cline{2-11}
&$\mathit{Tsk^4_2}$                  &[50,50]                            &5                          &0                          &50
&3                              &[1.2,1.9]  &-          &$Msg_3$    &$Msg_1$    \\
\cline{2-11}
&$\mathit{Tsk^4_3}$                  &[50,50]                            &25                          &0                          &50
&4                              &[0.1,0.2]  &-          &-          &$Msg_4$    \\
\cline{2-11}
&$\mathit{Tsk^4_4}$                  &[100,100]                          &11                          &0                          &100
&5                              &[0.7,1.1]  &-          &-          &-          \\
\cline{2-11}
&$\mathit{Tsk^4_5}$                  &[200,200]                          &13                          &0                          &200
&6                              &[3.7,5.8]  &-          &-          &-          \\
\hline\hline
\multirow{6}{*}{$P_5$}
&$\mathit{Tsk^5_1}$                  &[50,50]                            &0                          &0.3                          &50
&1                              &[0.7,1.1]  &-          &-          &$Msg_1$    \\
\cline{2-11}
&$\mathit{Tsk^5_2}$                  &[50,50]                            &2                          &0                          &50
&2                              &[1.2,1.9]  &-          &$Msg_4$    &$Msg_2$          \\
\cline{2-11}
&\multirow{2}{*}{$\mathit{Tsk^5_3}$} &\multirow{2}{*}{[200,200]}         &\multirow{2}{*}{0}         &\multirow{2}{*}{0}         &\multirow{2}{*}{200}            &\multirow{2}{*}{3}             &[0.4,0.6]  &-          &-          &-          \\\cline{8-11}
&&&&&&                          &[0.2,0.3]  &$Mux^5_1$  &-          &-          \\
\cline{2-11}
&\multirow{2}{*}{$\mathit{Tsk^5_4}$} &\multirow{2}{*}{[200,$\infty$)}    &\multirow{2}{*}{14}        &\multirow{2}{*}{0}         &\multirow{2}{*}{200}            &\multirow{2}{*}{4}             &[1.4,2.2]  &-          &-          &-          \\\cline{8-11}
&&&&&&                          &[0.1,0.2]  &$Mux^5_1$  &-          &-          \\
\hline\hline
\end{tabular}
\end{table}

\section{AFDX Configuration}

\begin{multicols}{2}
The AFDX configuration in Table \ref{tab:afdxconfig} is based on the case of \cite{gutierrez2014holistic}. There are four message types $\mathit{Msg_i},i=\{1,2,3,4\}$, each of which is allocated to a separate VL with the same subscript shown in column ``VL''. The column ``Length'' indicates the length of a message sent from an ARINC-653 partition. For any VL in the configuration, the columns ``BAG'' and ``$L_{max}$'' denote its Bandwidth Allocation Gap and Maximum packet Length respectively. The source and destination partition(s) are given in the columns ``Source'' and ``Destination'' respectively.
\end{multicols}

\begin{table}[!h]
\centering
\caption{The AFDX Configuration in the Case Study (Times in Milliseconds and Sizes in Bytes)}
\label{tab:afdxconfig}
\begin{tabular}{c|p{1cm}p{0.8cm}p{1cm}p{1cm}p{1cm}p{1.8cm}}
\hline\hline
\ Message       &Length         &VL         &BAG        &$L_{max}$      &Source         &Destinations       \\
\hline
$Msg_1$         &306            &$V_1$      &8          &200            &$P_1$          &$P_3$,$P_4$,$P_5$  \\
$Msg_2$         &953            &$V_2$      &16         &1000           &$P_2$          &$P_3$,$P_5$        \\
$Msg_3$         &453            &$V_3$      &32         &500            &$P_4$          &$P_3$              \\
$Msg_4$         &153            &$V_4$      &32         &200            &$P_5$          &$P_4$              \\
\hline\hline
\end{tabular}
\end{table}

\end{document}